\documentclass[10pt,twocolumn,twoside]{JCNtran}

\usepackage{amsmath,amssymb,amsfonts,graphicx,tikz}

\usepackage[dvips]{epsfig}
\usepackage[latin1]{inputenc}
\usepackage[T1]{fontenc}

\def\BibTeX{{\rm B\kern-.05em{\sc i\kern-.025em b}\kern-.08em
    T\kern-.1667em\lower.7ex\hbox{E}\kern-.125emX}}

\DeclareMathOperator{\GF}{GF}
\renewcommand{\vec}[1]{\ensuremath{\mathbf{#1}}}
\newcommand{\stxt}[1]{\ensuremath{_{\mathrm{#1}}}}
\newcommand{\utxt}[1]{\ensuremath{^{\mathrm{#1}}}}

\begin{document}
\bibliographystyle{ieeetr}
\title{Challenges and some new directions in channel coding}

\author{
Erdal Ar{\i}kan, Najeeb ul Hassan, Michael Lentmaier, Guido Montorsi and Jossy Sayir
\thanks{This work was supported by the FP7 Network of Excellence
NEWCOM\# under grant agreement~318306.}
\thanks{Erdal Ar{\i}kan is with Bilkent University, Ankara, Turkey.}
\thanks{Najeeb ul Hassan is with TU Dresden, Germany.}
\thanks{Michael Lentmaier is with Lund University, Sweden.}
\thanks{Guido Montorsi is with Politecnico di Torino, Italy.}
\thanks{Jossy Sayir is with the University of Cambridge, UK, email: j.sayir@ieee.org}
}
 \maketitle

\begin{abstract}
Three areas of ongoing research in channel coding are surveyed,
and recent developments are presented in each area: spatially
coupled Low-Density Parity-Check (LDPC) codes, non-binary LDPC
codes, and polar coding.
\end{abstract}


\section{Introduction}

The history of channel coding began hand in hand
with Shannon's information theory \cite{shannon1948}.
Following on the pioneering work of
Golay \cite{golay1949} and Hamming \cite{hamming1950},
the majority of linear codes developed in the early ages of coding theory
were ``error correction'' codes in the sense that their aim is
to correct errors made by the channel. The channel was
universally assumed to be a Binary Symmetric Channel (BSC).
The study of error correction codes culminated with the invention of
Reed-Solomon codes \cite{reedsolomon1960}
in 1960, which are Maximum Distance Separable (MDS) over non-binary fields
and hence are guaranteed to correct or detect the largest number of errors 
possible for a given code length and dimension. 

In parallel to the evolution of linear block codes, the invention
of convolutional codes by Peter Elias in 1955 
\cite{elias1955}
lead to a different 
approach and to the invention of trellis-based decoding methods
such as the Viterbi algorithm 
\cite{viterbi1967,forney1973}
and the BCJR algorithm \cite{bcjr}.
Both of these algorithms
can be easily adapted to any channel and hence generalise the concept
of error correction to general channels that cannot be described
simply in terms of probability of error.
We now speak of ``channel coding'' rather than ``error correction coding''.
Further progress in channel coding was made by Gottfried
Ungerboeck \cite{ungerboeck1982}
by linking coding to modulation for convolutional
codes. 

In 1993, Claude Berrou and co-authors shocked the coding research
community in \cite{berrou1993}
by designing a coding system known as ``turbo codes'' that
achieved a quantum leap in the performance of codes over general
channels. They obtained very good error performance within a small margin
of the channel capacity, something that had been thought impossible
with practical systems and moderate complexity by most coding theorists.
Yet Berrou's approach achieved this in an eminently implementable system
and with linear decoding complexity. In the subsequent scramble to explain 
the theory behind this puzzling performance, a method originally 
developed by Robert Gallager in his PhD thesis \cite{gallager-thesis},
known as Low-Density Parity-Check (LDPC) coding was rediscovered
in \cite{mackay1999}
and shown to have comparable properties. Both these methods
have become the workhorses of modern communication standards,
with arguments about the technical advantages of one over the
other mostly obscured by business and standardization interests of the
argumenter. What is clear and undisputed is that LDPC codes are
easier to explain and analyse and hence should probably take
precedence over turbo codes in teaching.
It is nowadays well-known that both LDPC codes and turbo codes can be
viewed as sparse codes on graphs. As a consequence they share a lot of
properties, and any construction or analysis method that can be
applied to one of them can usually be replicated for the
other. 
Some
technical differences between LDPC or turbo codes may tilt the balance
towards one or the other in specific applications.

We could conclude this history of coding here and bury the topic
into dusty textbooks, sending it the same way as classical Newtonian
mechanics\footnote{Apologies to mechanics researchers for the
seemingly disparaging remark. In fact, we are aware that classical
mechanics is an ongoing and modern research topic as evidenced
by many journals and conferences, just as coding theory is.}
and other topics made obsolete by quantum leaps in research.
Many coding researchers nowadays are confronted with the recurrent
``Coding is dead'' motto \cite{massey1974}
of experts claiming that, now that 
capacity is achieved, there is nothing further to be researched
in the field. In fact, as this paper will contribute to showing, 
coding is still an ongoing and very active topic of research
with advances and innovations to address important and 
practical unsolved problems.

Current hurdles in the applicability of modern coding techniques
can be classified in two categories:
\begin{description}
\item[Complexity] While turbo and LDPC codes have brought 
capacity-approaching performance within reach of implementable
systems, implementable does not necessarily mean practical.
The complexity of codes that perform well under practical
constraints such as limited decoding delay and high spectral
efficiency is still a major hurdle for low power implementations
in integrated circuits. There is a serious need for new methods
that simplify code design, construction, storage, and decoder
implementation.
\item[New applications] Turbo and LDPC codes can be seen to
``solve'' the capacity problem for elementary point-to-point 
channels. Recent years have seen advances in information
theory for many multi-user channels such as the multiple access,
broadcast, relay and interference channels. As communication
standards become more ambitious in exploiting the available
physical resources such as spectrum and geographical reach,
there is a push to switch from interference
limited parallel point-to-point protocols to true multi-user
processing with joint encoding and/or decoding. There is 
a need for coding methods that can do this efficiently for
all of the scenarios described. Furthermore, theory has gone
further than pure communications by expanding to distributed
compression and joint source/channel coding, distributed
storage, network coding, and quantum channels and protocols.
All of these new theories come with
their own requirements and constraints for coding, and hence
coding research is far from dead when it comes to these
new applications.
\end{description}
The paper will present three areas of ongoing research
in coding, all of which have some degree of relevance to
the two challenges described.

In Section~\ref{sec:SCLDPC}, we will address spatially coupled LDPC codes, which
have a structure akin convolutional codes. For spatially coupled codes
the asymptotic performance of an iterative decoder is improved to that
of an optimal decoder, which opens the way for new degrees of freedom
in the code design. For example, it is possible to achieve capacity
universally for a large class of channels with simple regular SC-LDPC
codes where irregular LDPC codes would require careful individual
optimizations of their degree profiles. We will discuss the design of
SC-LDPC codes for flexible rates, efficient window decoding techniques
for reduced complexity and latency, and the robustness of their
decoding for mobile radio channels.
In Section~\ref{sec:non-binary},
we will address non-binary LDPC and related codes. These
are codes over higher order alphabets that can, for example,
be mapped directly onto a modulation alphabet, making them
interesting for high spectral efficiency applications. While
these have been known for a while, the complexity of decoding
has made them unsuited for most practical applications. In this
section, we will discuss research advances in low complexity
decoding and also present a class of LDPC codes with an associated 
novel decoding algorithm known as Analog Digital Belief Propagation (ADBP)
whose complexity does not increase with alphabet size and hence
constitutes a promising development for very high spectral
efficiency communications. Finally, in Section~\ref{sec:polar},
we will introduce Polar coding, a new technique introduced
in \cite{ArikanIT2009} based on a phenomenon known as
channel polarization, that has the flexibility and versatility
to be an interesting contender for many novel application scenarios.

\section {Spatially Coupled LDPC Codes} \label{sec:SCLDPC}

The roots of low-density parity-check (LDPC) codes \cite{gallager-thesis} trace back to the concept of random coding. It can be shown that a randomly generated code decoded with an optimal decoder exhibits very good performance with high probability. However, such a decoder is infeasible in practice because the complexity will increase exponentially with the code length. The 
groundbreaking 
idea of Gallager was to slightly change the random ensemble in such a way that the codes can be decoded efficiently by an iterative algorithm, now known as belief propagation (BP) decoding. His LDPC codes were defined by sparse parity-check matrices $\vec{H}$ that contained a fixed number of $K$ and $J$ non-zero values in every row and column, respectively, known as {\em regular} LDPC codes. Gallager was able to show that the minimum distance of typical codes of the ensemble grows linearly with the block length, which guarantees that very strong codes can be constructed if large blocks are allowed. The complexity per decoded bit, on the other hand, is independent of the length if the number of decoding iterations is fixed.

The asymptotic performance of an iterative decoder can be analyzed by tracking the probability distributions of messages that are exchanged between nodes in the Tanner graph ({\em density evolution}) \cite{RU01}. The worst channel parameter for which the decoding error probability converges to zero is called the {\em BP threshold}. The BP thresholds of turbo codes are actually better than those of the original regular LDPC codes of Gallager. A better BP threshold is obtained by allowing the nodes in the Tanner graph to have different degrees \cite{RU01}. By optimizing the degrees of the resulting {\em irregular} LDPC code ensembles it is possible to push the BP thresholds towards capacity. However, this requires a large fraction of low-degree variable nodes, which leads to higher error floors at large SNRs. As a consequence of the degree optimization, the capacity achieving sequences of irregular LDPC codes do no longer show a linear growth of the minimum distance.

\begin{figure}
	\begin{center}
	\includegraphics[width=0.8\linewidth]{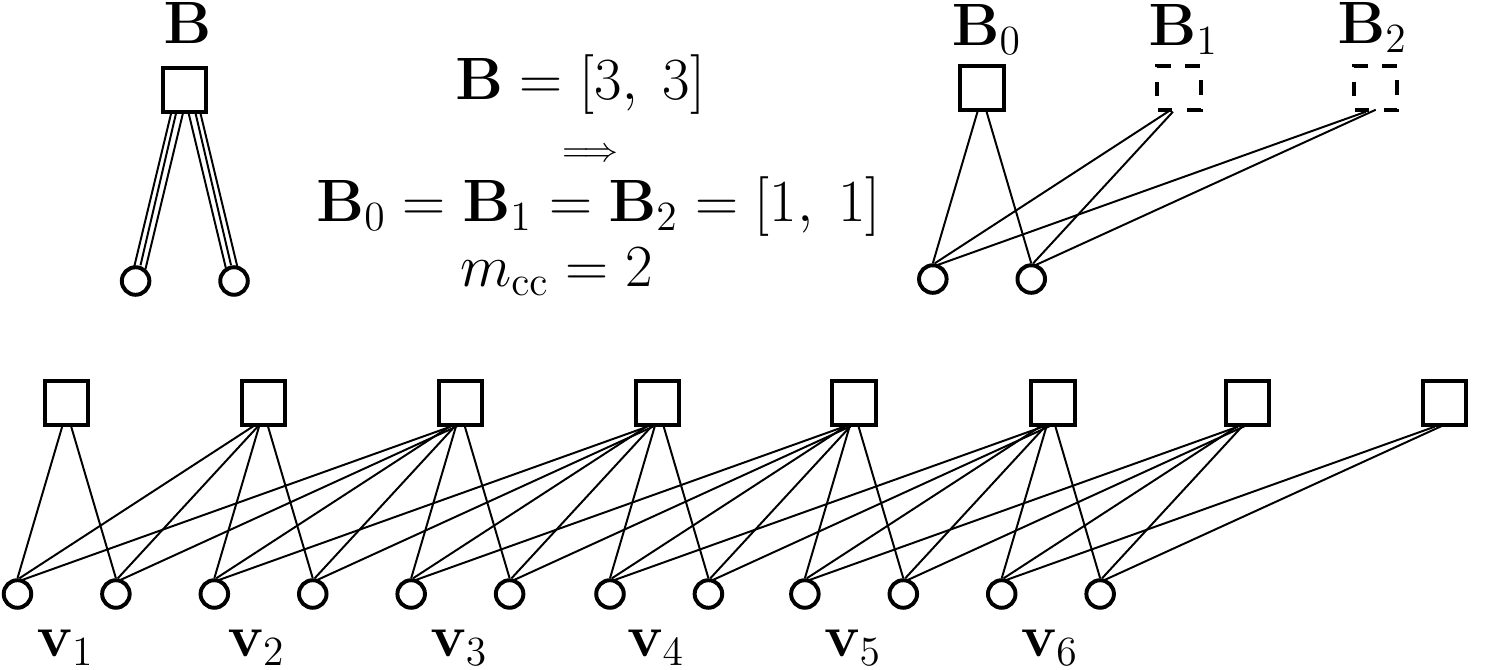}
	\caption{Illustration of edge spreading: the protograph of a (3,6)-regular block code represented by a base matrix $\vec{B}$ is repeated $L=6$ times and the edges are spread over time according to the component base matrices  $\vec{B}_0$, $\vec{B}_1$, and $\vec{B}_2$, resulting in a terminated LDPCC code.}
	\label{fig:EdgeSpreading}
\end{center}
\end{figure}

LDPC convolutional codes were invented by Jim\'{e}nez Feltstr\"{o}m 
and Zigangirov in \cite{JZ99}. Like LDPC block codes, they are defined
by sparse parity-check matrices, which can be infinite but have a
band-diagonal structure like the generator matrices of classical
convolutional codes.  When the parity-check matrix is composed of
individual permutation matrices, the structure of an LDPC code ensemble
can be described by a {\em protograph} \cite{Tho03} (a prototype
graph) and its corresponding base matrix $\vec{B}$. 
The graph of an LDPC convolutional code can be obtained by
starting from a sequence of $L$ independent protographs of an LDPC
block code, which are then interconnected by spreading the edges over
blocks of different time instants \cite{LFZC09}. The maximum width of
this {\em edge spreading} determines the memory, $m\stxt{cc}$, of the
resulting chain of length $L$ that defines the LDPC convolutional code
ensemble. Since the blocks of the original protograph codes are
coupled together by this procedure, LDPC convolutional codes are also
called {\em spatially coupled} LDPC codes
(SC-LDPC). Figure~\ref{fig:EdgeSpreading} shows an illustration of the
edge spreading procedure.

A BP threshold analysis of LDPC convolutional codes shows that the performance of the iterative decoder is improved significantly by spatial coupling. In fact, the results in \cite{LSCZ10} show that asymptotically, as $L$ tends to infinity, the BP threshold is boosted to that of the optimal maximum a posteriori (MAP) decoder. Stimulated by these findings, Kudekar, Richardson and Urbanke developed an analytical proof of this {\em threshold saturation} phenomenon \cite{KRU11}\cite{KRU12}. More recently, potential functions have been identified as a powerful tool for characterizing the connection between MAP thresholds and BP thresholds \cite{KYMP12}. All these approaches make use of the {\em area theorem} \cite{MU05} in order to derive bounds on the MAP threshold and prove threshold saturation for spatially coupled codes. Since the MAP thresholds of regular LDPC ensembles with increasing node degrees are known to converge to capacity, it follows that spatial coupling provides a new way of provably achieving capacity with low-complexity iterative BP decoding --- not only for the BEC but also for the AWGN channel. Furthermore, the spatially coupled code ensembles inherit from the uncoupled counterparts, the linearly increasing minimum distance property \cite{MLC10}. This combination of capacity achieving thresholds with low complexity decoding and linearly increasing distance is quite unique and has attracted a lot of interest in the research community.

\begin{figure}
	\begin{center}
	\includegraphics[width=0.8\linewidth]{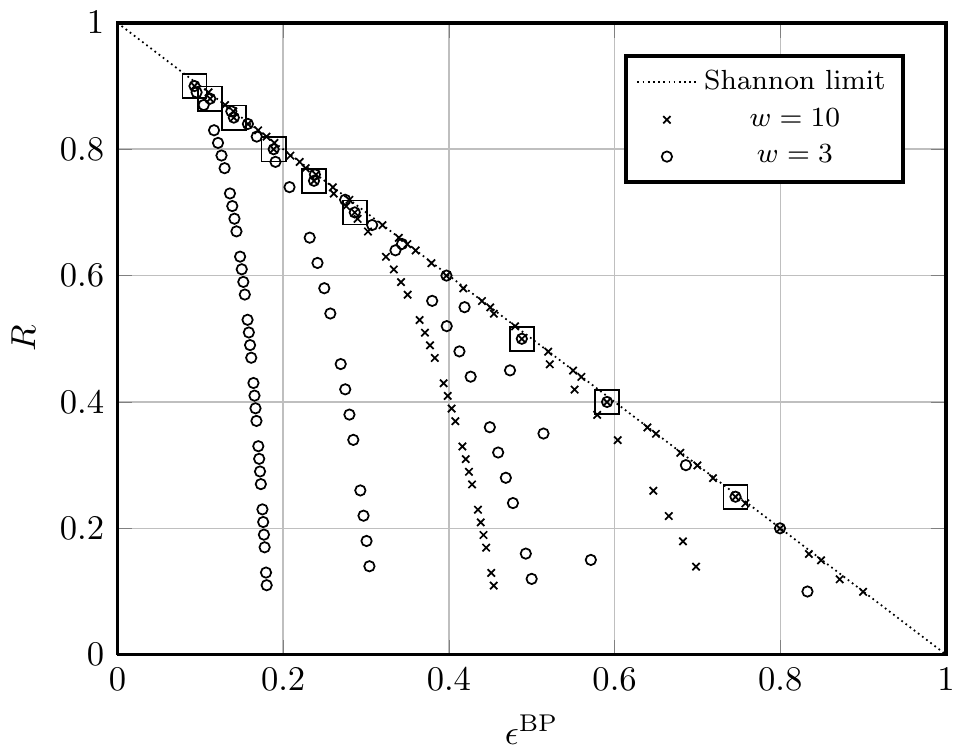}
	\caption{Density evolution thresholds $\epsilon\utxt{BP}$ for
          $(J,K)$-regular SC-LDPC ensembles in comparison with the
          Shannon limit $\epsilon\utxt{Sh}$. 
The coupling width $w$ is equal  to $m\stxt{cc}+1$. For a given rate $R=1-J/K$, the
          smallest pair of values $J$ and $K$ are chosen under the
          condition that $J \geq 3$.  The ensembles
with minimum variable node degree $J=3$ are highlighted with squares.} 
	\label{fig:nearlyReg}
\end{center}
\end{figure}

The capacity achieving property of regular SC-LDPC codes raises the question whether irregularity is still needed at all. In principle, it is possible for any arbitrary rational rate to construct regular codes that guarantee a vanishing gap to capacity with BP decoding. On the other hand, for some specific code rates, the required node degrees and hence the decoding complexity increase drastically. But even if we neglect the complexity, there exists another problem of practical significance that so far has not received much attention in the literature: for large node degrees $J$ and $K$ the threshold saturation effect will only occur for larger values of the coupling parameter $m\stxt{cc}$, as illustrated for the BEC in Fig.~\ref{fig:nearlyReg} \cite{NLF14}. 
We can see that for a given coupling width $w=m\stxt{cc}+1$,  the gap to capacity becomes small only for certain code rates $R$, and it turns out that these rates correspond to the ensembles for which the variable node degree $J$ is small.

Motivated by 
this observation, in \cite{NLF14} some
nearly-regular SC-LDPC code ensembles where introduced, which are
built upon the mixture of two favorable regular codes of same variable
node degree. The key is to allow for a slight irregularity in the code
graph to add a degree of freedom that can be used for supporting
arbitrary rational rates as accurately as needed while keeping the
check and variable degrees as low as possible. These codes exhibit
performance close to the Shannon limit for all rates in the
rate interval considered,
while having a decoder complexity as low as for the best regular
codes. The exclusion of variable nodes of degree two in the
construction ensures that the minimum distance of the proposed
ensembles increases linearly with the block length, i.e., the codes
are asymptotically good.

\subsection{Efficient Decoding of Spatially Coupled Codes}

\begin{figure}[t]
 \centering
\begin{tikzpicture}
\foreach \x in {0,...,6}
{
 \draw (1+\x-.1,1.9) rectangle node (r\x) {} (1+\x+.1,2.1);
}
\foreach \x in {0,...,3}
{
 \draw [fill=green] (.9+\x/2,1) circle [radius=.1] node (c\x) {} ;
}
\foreach \x in {4,5}
{
 \draw [fill=red] (.9+\x/2,1) circle [radius=.1] node (c\x) {} ;
}
\foreach \x in {6,...,13}
{
 \draw (.9+\x/2,1) circle [radius=.1] node (c\x) {} ;
}
\draw [<->] (.7,2.5) -- node [above] {$m_{cc}=2$} (2.6,2.5);
\draw [<->] (2.65,2.5) -- node [above] {$W=4$} (6.6,2.5);
\draw [dashed] (.7,.5) rectangle (2.6,2.3);
\draw (2.65,.5) rectangle (6.6,2.3);
\node [above] at (3.15,.5) {\scriptsize $w=1$};
\node [above] at (4.15,.5) {\scriptsize $w=2$};
\node [above] at (5.15,.5) {\scriptsize $w=3$};
\node [above] at (6.15,.5) {\scriptsize $w=4$};
\node at (1.1,0.3) {$y_{t-2}$};
\node at (2.1,0.3) {$y_{t-1}$};
\node at (3.1,0.3) {$y_t$};
\node at (4.1,0.3) {$y_{t+1}$};
\node at (5.1,0.3) {$y_{t+2}$};
\node at (6.1,0.3) {$y_{t+3}$};
\node at (7.1,0.3) {$y_{t+4}$};
\node at (0,1.5) {\ldots};
\node at (8.3,1.5) {\ldots};
\draw [draw=gray] (-.1,1) -- (r0);
\draw [draw=gray] (-.1,1) -- (r1);
\draw [draw=gray] (.4,1) -- (r0);
\draw [draw=gray] (.4,1) -- (r1);
\draw [draw=gray] (c0) -- (r0);
\draw [draw=gray] (c0) -- (r1);
\draw [dashed] (c0) -- (r2);
\draw [draw=gray] (c1) -- (r0);
\draw [draw=gray] (c1) -- (r1);
\draw [dashed] (c1) -- (r2);
\draw [draw=gray] (c2) -- (r1);
\draw [dashed] (c2) -- (r2);
\draw [dashed] (c2) -- (r3);
\draw [draw=gray] (c3) -- (r1);
\draw [dashed] (c3) -- (r2);
\draw [dashed] (c3) -- (r3);
\draw (c4) -- (r2);
\draw (c4) -- (r3);
\draw (c4) -- (r4);
\draw (c5) -- (r2);
\draw (c5) -- (r3);
\draw (c5) -- (r4);
\draw (c6) -- (r3);
\draw (c6) -- (r4);
\draw (c6) -- (r5);
\draw (c7) -- (r3);
\draw (c7) -- (r4);
\draw (c7) -- (r5);
\draw (c8) -- (r4);
\draw (c8) -- (r5);
\draw [draw=gray] (c8) -- (r6);
\draw (c9) -- (r4);
\draw (c9) -- (r5);
\draw [draw=gray] (c9) -- (r6);
\draw (c10) -- (r5);
\draw [draw=gray] (c10) -- (r6);
\draw [draw=gray] (c10) -- (8,2);
\draw (c11) -- (r5);
\draw [draw=gray] (c11) -- (r6);
\draw [draw=gray] (c11) -- (8,2);
\draw [draw=gray] (c12) -- (r6);
\draw [draw=gray] (c12) -- (8,2);
\draw [draw=gray] (c12) -- (8,1.4);
\draw [draw=gray] (c13) -- (r6);
\draw [draw=gray] (c13) -- (8,2);
\draw [draw=gray] (c13) -- (8,1.3);
\end{tikzpicture}
    \caption{\label{fig:WD} Window decoder of size $W=4$ at time $t$. The green variable nodes represent decoded blocks and the red variable nodes ($\vec{y}_t$) are the target block within the current window. The dashed lines represent the read access to the $m\stxt{cc}$ previously decoded blocks.}
\end{figure}
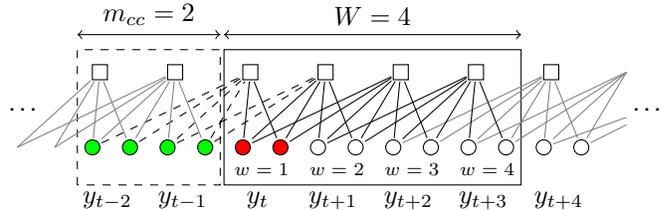

In order to achieve the MAP threshold, the number $L$ of coupled code blocks should be sufficiently large for reducing the rate loss due to termination of the chain. But running the BP decoder over the complete chain of length $L$ would then result in a large latency and decoding complexity and hence is not feasible in practical scenarios.
However, thanks to the limited width of the non-zero region around the diagonal, SC-LDPC codes can be decoded in a continuous fashion using a sliding window decoder \cite{Iyengar2012} of size $W$ ($W \ll L$). 
As a result, decoding latency and decoding complexity become independent of $L$. Moreover, the storage requirements for the decoder are reduced by a factor of $L/W$ compared to a non-windowed decoder. An example of the window decoder of size $W=4$ is given in Fig.~\ref{fig:WD}. 

It has been shown in \cite{HLF+12} that for equal structural latency,
SC-LDPC codes under window decoding 
 outperform LDPC codes for short to long
latency values and outperform convolutional codes from medium to long
latency values. For applications requiring very short latency, Viterbi
decoded convolutional codes were still found to be the optimal choice
\cite{HH09}\cite{MCFF10}\cite{HLF+12}. Note that only structural
latency was considered in all these comparisons which is defined as
the number of bits required before decoding can start. It therefore
can be concluded that for low transmission rate applications (in the
range of bit/seconds), convolutional codes with moderate constraint
length are favorable since the delay in filling the decoder buffer
dominates the overall latency.  Whereas, for applications with
transmission rates in excess of several Gigabit/seconds, e.g., short
range communication, medium to large structural latency is tolerable
and strong codes such as SC-LDPC codes provide gain in performance
compared to the conventional convolutional codes.  Another advantage
of using a window decoder is the flexibility in terms of decoding
latency at the decoder. Since the window size $W$ is a decoder
parameter, it can be varied without changing the code, providing a
flexible trade-off between performance and latency \cite{HLF+12}.

In BP decoding, messages are passed between the check and variable
nodes according to a parallel (flooding) or serial (on-demand)
rule
\cite{SPL09}. 
In both schedules, all the nodes in the graph are typically
updated at every decoding iteration (\emph{uniform schedules}).  For
both LDPC and SC-LDPC, a uniform serial decoding schedule results in a
factor of two in complexity reduction when applied over the complete
length of the code \cite{SPL09}. However, this gain in complexity
reduction reduces to only $20\%$ when uniform serial schedules are
applied within a decoding window \cite{HPLFC12}\cite{HPLFC13}.  In
order to reduce the decoding complexity for window decoding,
non-uniform window decoding schedules has been introduced in
\cite{HPLFC12}\cite{HPLFC13}, which result in $50\%$ reduction in
complexity compared to uniform decoding schedules.  The reduction in
decoding complexity can be achieved by avoiding unnecessary updates of
nodes not directly connected to the first position in the window. Only
nodes that show improvement based on their BER compared to the
previous iteration are updated in the next iteration.

\subsection{Performance over Mobile Radio Channels}

One of the most remarkable features of spatially coupled codes is their universality property, which means that a single code construction performs well for a large variety of channel conditions. For discrete-input memoryless symmetric channels the universality of SC-LDPC codes has been shown in \cite{KRU12}. In this section we consider the block-fading channel and demonstrate that  SC-LDPC codes show a remarkable performance on this class of channels.

\begin{figure}[t]
 \centering
    \includegraphics[width=0.7\linewidth]{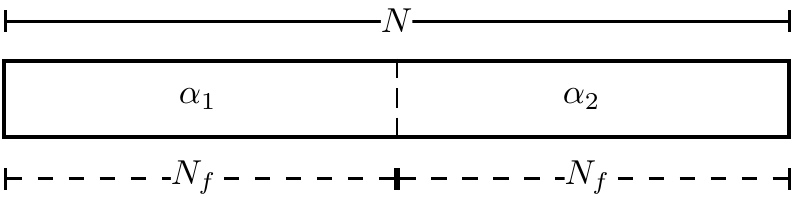}
    \caption{\label{fig:BFmodel} Illustration of block-fading channel for a codeword of length $N$ and $F=2$.}
\end{figure}

The block-fading channel was introduced in \cite{OSW94} to model the mobile-radio environment. This model is useful because the channel coherence time in many cases is much longer than one symbol duration and several symbols are affected by the same fading coefficient. 
The coded information is transmitted over a finite number of fading blocks to provide diversity. An example where a codeword of length $N$ spreads across $F=2$ fading realizations is shown in Fig.~\ref{fig:BFmodel}.
In general, when dealing with block-fading channels, two strategies can be adopted: coding with block interleaving or coding with memory \cite{BCT00}. Spatially-coupled codes, with their convolutional structure among LDPC codes, are expected to be a nice example of the second strategy. 

The block-fading channel is characterized by an \emph{outage
probability}, which serves as a lower bound on the word error
probability for any code decoded using a maximum likelihood decoder.
In terms of density evolution, the {density evolution outage} (DEO) 
is the event when the bit error probability does not converge to zero for a
fixed value of SNR after a finite or an infinite number of decoding
iterations are performed \cite{BGBZ10}. The probability of density
evolution outage, for a fixed value of SNR, can then be calculated
using a Monte Carlo method considering significant number of fading
coefficients.

\begin{figure}[t]
 \centering
    \includegraphics[width=0.9\linewidth]{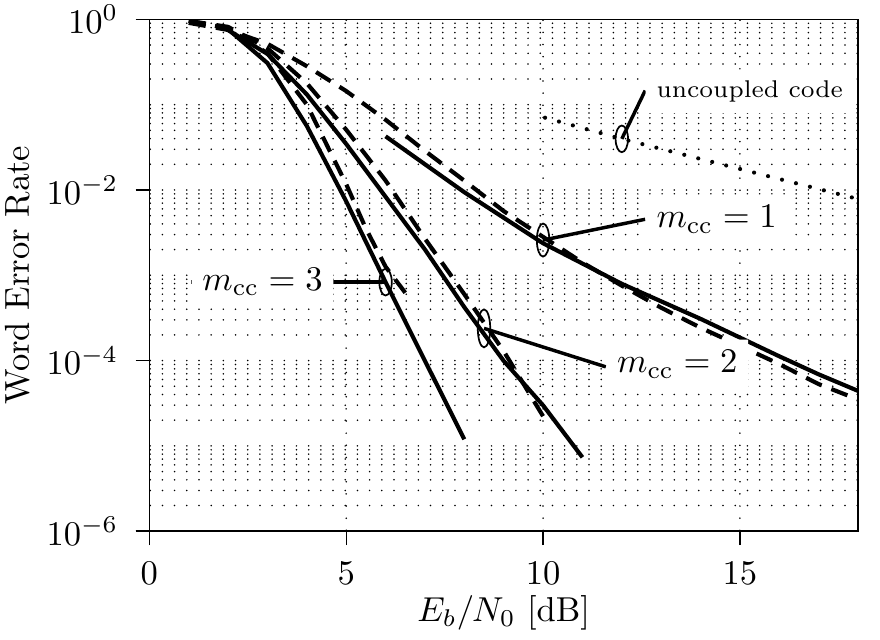}
    \caption{\label{fig:blFad} Density evolution outage for SC-LDPC codes with memory 0,1,2 and 3. The bold lines represent the DEO and dashed lines represent the simulation results when a code with $N = 200$, $L = 100$, is decoded using a window decoder, $F = 2$.}
\end{figure}

Since the memory of the code plays an important role to exploit code diversity, we consider SC-LDPC codes with increasing memory from $0$ to $3$. 
The diversity of the code, which is defined as the slope of the WER curve, is calculated numerically from the DEO curves presented in Fig.~\ref{fig:blFad}. 
For uncoupled LDPC codes, the diversity is limited to $d=1.3$ (see
dotted line in Fig.~\ref{fig:blFad}). This case can be interpreted as
an SC-LDPC code with  $m\stxt{cc}=0$. If we now increase the coupling
parameter to $1$, $2$ and $3$, then the diversity of SC-LDPC codes
increases to $3$, $6$ and $10$, respectively \cite{HLAF14}. 
The figure also shows the simulation results (dashed lines) for finite length codes when the length of each individual coupled code block is $N=200$. The simulation results match closely with the calculated DEO bounds.

An alternative approach to codes with memory is taken by the \emph{root-LDPC} codes \cite{BGBZ10} with a special check node structure called \textit{rootcheck}. Full diversity ($d=F=1/R$) is provided to the systematic information bits only by connecting only one information bit to every rootcheck.
However, designing root-LDPC codes with diversity order greater than $2$ requires codes with rate less than $R=1/2$. The special structure of the codes makes it a complicated task to generate good root-LDPC codes with high diversity (and thus low rate). 

Another key feature of SC-LDPC codes is its robustness against the variation in the channel.  
In case of root-LDPC codes, the parity-check matrix has to be designed for the specific channel parameter $F$ to provide a diversity of $d=F$ to the information bits. However for SC-LDPC codes, it can be shown that the code design for a specific value of $F$ is not required whereas the diversity order strongly depends on the memory of the code. 
This feature makes them very suitable for a wireless mobile environment.

\section{Non-Binary Codes and High Spectral Efficiency Codes}
\label{sec:non-binary}

Low-Density Parity-Check (LDPC) codes
were originally proposed by Gallager \cite{gallager-thesis}
and re-discovered by MacKay \& al. \cite{mackay1999} in the years after 
the invention of turbo codes \cite{berrou1993}. LDPC codes have been adopted
in several current standards, e.g., IEEE 802.11n Wi-Fi standard, DVB-S2, T2, and C2 digital
video broadcasting satellite, cable and terrestrian, 10GBase-T ethernet over twisted pairs,
G.hn/G.9960 home networking over power lines. Together with turbo codes, they are the 
modern coding technique of choice when it comes to designing communication systems that
approach the theoretical limits of physical transmission media in terms of data rate, 
transmission power, geographical reach and reliability. 

All LDPC codes in current standards are binary codes. LDPC codes over non-binary
alphabets were mentioned in \cite{gallager-thesis} and fully described in \cite{davey1998}.
They offer two practical advantages and one major disadvantage with respect to 
binary codes: 
\begin{itemize}
\item Advantage 1: encoding directly over the $q$-ary alphabet corresponding to the signal 
constellation used for modulation saves the mapping and de-mapping operations
needed to transfer between binary coding alphabet and non-binary modulation
signal space. Furthermore, the de-mapping operation is costly in terms of 
complexity and introduces a loss of sufficient statistic and a resulting performance
loss that can only be partially countered by proper choice of the mapping,
or fully recovered by costly iterations over the de-mapper and the decoder.
With non-binary codes, there is no mapping and no loss of efficiency through
de-mapping as the input messages to the decoder are a sufficient statistic
for the transmitted symbols, making non-binary LDPC codes a tempting proposition
for high spectral efficiency coding over higher order constellations.
\item Advantage 2: non-binary LDPC codes tend to exhibit less of a performance loss when
the block length is shortened to accommodate delay constraints, as compared
to binary codes. 
\item Disadvantage: the decoding complexity of LDPC codes increases with the 
alphabet size.
\end{itemize}
The complexity issue has been addressed in a number of refinements of the non-binary
LDPC iterative decoding algorithm. The plain description of the decoder requires 
convolutions of $q$-ary distribution-valued messages in every constraint node of the
associated factor graph. A first and appealing improvement \cite{davey1998} is obtained by 
switching to the frequency domain where convolutions become multiplications. This
involves taking the $q$ point discrete Fourier transform (DFT) if $q$ is a 
prime number, or, for the more practical case where $q$ is a power of two $q=2^m$,
taking the $q$ point Walsh-Hadamard transform (WHT). This step reduces the 
constraint node complexity from $q^2$ to $q\log q$ by evaluating the appropriate
transform in its ``fast'' butterfly-based implementation, i.e., Fast Fourier transform (FFT)
for the DFT and Fast Hadamard transform (FHT) for the WHT.

While this first improvement is significant, the resulting complexity is still much
higher than that of the equivalent binary decoder. The currently least complex methods known
for decoding non-binary LDPC codes are various realizations of the Extended Min-Sum (EMS)
\cite{declercq2006} algorithm. In this method, convolutions are evaluated directly
in the time domain but messages are first truncated to their most significant components,
and convolutions are evaluated on the truncated alphabets, resulting in a significant 
complexity reduction with respect to the $q^2$ operations needed for a full convolution.
While the principle of the algorithm is easy enough to describe as we just did, in fact
its implementation is quite subtle because of the need to remember which symbols are
retained in the truncated alphabet for each message and which configurations of input
symbols map to which output symbols in a convolution. Many technical improvements
of the EMS can be achieved by hardware-aware implementation of the convolution 
operations, e.g., \cite{voicila2010}, \cite{boutillon2010}.

In this section, we discuss two current research areas related to non-binary
codes. First, we will look at frequency-domain methods that operate on
truncated messages. The aim here is to achieve a fairer comparison
of complexity between the EMS and frequency-domain methods, since much
of the gain of the EMS is achieved through message truncation, but in
complexity comparisons it is evaluated alongside frequency domain
decoders operating on full message sets. In the second part of this
section, we will look at a novel non-binary code construction operating
over rings rather than fields, with a decoding algorithm known as
Analog Digital Belief Proapagation (APBP) \cite{montorsi2012analog}.
This promising new approach has
the merit that its complexity does not increase with the alphabet
size, in contrast to regular belief propagation for LDPC codes over $q$-ary fields,
making it an appealing proposition for very high spectral efficiency
communications.

\subsection{Frequency domain decoding with truncated messages}

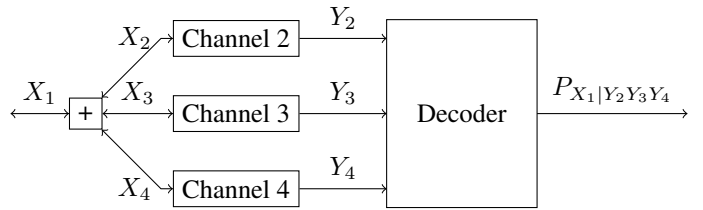
\begin{figure}[h]
\centering
\begin{tikzpicture}
\tikzstyle{rect}=[rectangle, draw=black, minimum size=3mm];
\node (plus) at (1,1) [rect] {+};
\node (ch2) at (3,2) [rect] {Channel 2};
\node (ch3) at (3,1) [rect] {Channel 3};
\node (ch4) at (3,0) [rect] {Channel 4};
\draw [<->] (plus) -- (2,2) node [left] {$X_2$} -- (ch2);
\draw [<->] (plus) -- node [above] {$X_3$} (ch3);
\draw [<->] (plus) -- (2,0) node [left] {$X_4$} -- (ch4);
\draw (5,-.25) rectangle node {Decoder} (7,2.25);
\draw [->] (ch2) -- node [above] {$Y_2$} (5,2);
\draw [->] (ch3) -- node [above] {$Y_3$} (5,1);
\draw [->] (ch4) -- node [above] {$Y_4$} (5,0);
\draw [->] (7,1) -- node [above] {$P_{X_1|Y_2Y_3Y_4}$} (9,1);
\draw [<->] (0,1) -- node [above] {$X_1$} (plus);
\end{tikzpicture}
\caption{Conceptual scenario for a degree 4 constraint node decoder}
\label{fig:constraint-decoder}
\end{figure}

The ideal constraint node operation of an LPDC decoder operating on 
a field $\mathcal{F}$ implements a Bayesian estimator for the
conceptual scenario illustrated in Figure~\ref{fig:constraint-decoder}.
The estimator provides the a-posteriori probability distribution
of code symbol $X_1$ given the observations $Y_2,Y_3$ and $Y_4$
of the code symbols $X_2,X_3$ and $X_4$, respectively, where the
sum of $X_1,X_2,X_3$ and $X_4$ is zero over $\mathcal{F}$.
Assuming that the input to the decoder is provided in terms
of a-posteriori probability distributions $P_{X_2|Y_2}$, $P_{X_3|Y_3}$
and $P_{X_4|Y_4}$, i.e., as distribution-valued messages, it follows
that the distribution $P_{X_1|Y_2Y_3Y_4}$ to be computed is a 
type of convolution of the input distributions. For example, if
$\mathcal{F} =\GF(\mbox{3})$ 
, i.e., the field of numbers $\{0,1,2\}$ using
arithmetic modulo 3, then the output probability that $X_1$ be zero
given $Y_2,Y_3$ and $Y_4$ is the sum of the probabilities all
configurations of $X_2,X_3$ and $X_4$ that sum to zero, i.e.,
0,0,0 or 0,1,2 or 0,2,1 or 1,0,2 or 1,1,1 or 1,2,0 or 2,0,1 or 2,1,0
or 2,2,2. This case results in a cyclic convolution of the three
distribution-valued input messages. Over the more commonly used
binary extension fields $\GF(2^m)$, where the sum is defined as
a bitwise sum, the corresponding operation is a componentwise 
cyclic convolution in multi-dimensional binary space.

\begin{figure}[h]
\centering
\begin{tikzpicture}
\draw (0,3) rectangle node (t1) {Transform} (2,4);
\draw (2.5,3) rectangle node (t2) {Transform} (4.5,4);
\draw (5,3) rectangle node (t3) {Transform} (7,4);
\draw (.5,2.5) circle [radius=.15] node (m1) {$\times$}; 
\draw (1,2) circle [radius=.15] node (m2) {$\times$}; 
\draw (1.5,1.5) circle [radius=.15] node (m3) {$\times$}; 
\draw (3,2.5) circle [radius=.15] node (m4) {$\times$}; 
\draw (3.5,2) circle [radius=.15] node (m5) {$\times$}; 
\draw (4,1.5) circle [radius=.15] node (m6) {$\times$}; 
\draw (0,0) rectangle node (it) {Transform$^{-1}$} (2,1);
\draw [->] (.5,4.5) -- (.5,4);
\draw [->] (1,4.5) -- (1,4);
\draw [->] (1.5,4.5) -- (1.5,4);
\draw [->] (3,4.5) -- (3,4);
\draw [->] (3.5,4.5) -- (3.5,4);
\draw [->] (4,4.5) -- (4,4);
\draw [->] (5.5,4.5) -- (5.5,4);
\draw [->] (6,4.5) -- (6,4);
\draw [->] (6.5,4.5) -- (6.5,4);
\node [above] at (1,4.5) {$P_{X_2|Y_2}$};
\node [above] at (3.5,4.5) {$P_{X_3|Y_3}$};
\node [above] at (6,4.5) {$P_{X_4|Y_4}$};
\draw [->] (.5,3) -- (.5,2.65);
\draw [->] (1,3) -- (1,2.15);
\draw [->] (1.5,3) -- (1.5,1.65);
\draw [->] (3,3) -- (3,2.65);
\draw [->] (3.5,3) -- (3.5,2.15);
\draw [->] (4,3) -- (4,1.65);
\draw [->] (.5,2.35) -- (.5,1);
\draw [->] (1,1.85) -- (1,1);
\draw [->] (1.5,1.35) -- (1.5,1);
\draw [->] (2.85,2.5) -- (1.6,2.5) arc (0:180:.1) -- (1.1,2.5) arc (0:180:.1) -- (.65,2.5);
\draw [->] (3.35,2) -- (1.6,2) arc (0:180:.1) -- (1.15,2);
\draw [->] (3.85,1.5) -- (1.65,1.5);
\draw [->] (5.5,3) -- (5.5,2.5) -- (4.1,2.5) arc (0:180:.1) -- (3.6,2.5) arc (0:180:.1) -- (3.15,2.5);
\draw [->] (6,3) -- (6,2) -- (4.1,2) arc (0:180:.1) -- (3.65,2);
\draw [->] (6.5,3) -- (6.5,1.5) -- (4.15,1.5);
\draw [->] (.5,0) -- (.5,-.5);
\draw [->] (1,0) -- (1,-.5);
\draw [->] (1.5,0) -- (1.5,-.5);
\node [below] at (1,-.5) {$P_{X_1|Y_2Y_3Y_4}$};
\end{tikzpicture}
\caption{Frequency domain convolution}
\label{fig:fconv}
\end{figure}
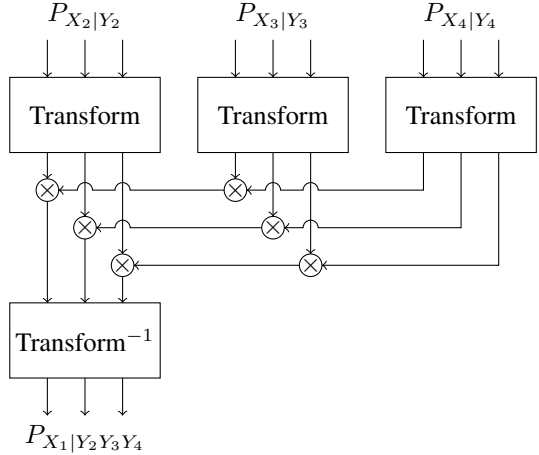

Convolution can be efficiently operated in the frequency domain. 
For a pure cyclic convolution such as the one illustrated over
$\GF(3)$, the transform required is the discrete Fourier transform (DFT).
The convolution of vectors in the time domain is equivalent
to the componentwise product of the corresponding vectors in the
transform domain. This process is illustrated in Figure~\ref{fig:fconv}.
For the more practically relevant binary extension fields $\GF(2^m)$, 
the same process applies but the transform required is the Walsh-Hadamard
transform (WHT).  

Both the DFT and the WHT can be operated efficiently using a
fast butterfly structure as the Fast Fourier transform (FFT) or the
Fast Hadamard Transform (FHT), requiring $q\log q$ operations 
where $q$ is the alphabet size of the code. In a typical non-binary LDPC 
decoder realization, these transforms despite their efficient 
implementation still use up over 90\% of the computing resources
and hence constitute the main hurdle for the practical implementability
of non-binary LDPC when compared to binary LDPC codes.
The approach of the EMS is to revert to time-domain convolutions
but operate them on reduced alphabet sizes $q'\ll q$ by truncating
each incoming distribution-valued message to its largest components.
The resulting algorithm is more difficult to operate than may at
first appear, because in such partial convolutions one needs to retain
which output values emerge from the mappings of the differing truncated
alphabets of each input message, so the implementation needs to 
perform operations in $\mathcal{F}$ in parallel to the convolution
operations over the probabilities. The complexity comparison becomes
a comparison between $q'^2$ and $q\log q$. For example, when operating
in $\GF(64)$, the complexity of the frequency domain based decoder
is on the order of $6\times 64=384$ operations per constraint node
per iteration, whereas the EMS with messages truncated to $q'=8$ is 
in the order of $8\times 8=64$ operations per constraint node per
iteration. An added benefit of performing convolutions in the time
domain is that one can operate in the logarithmic domain, replacing
products by max operations using the well established approach
that also underpins the min-sum method for decoding binary LDPC codes.

The comparison described above is not completely fair because
it fails to take into account that message truncation may also
be of benefit when operating in the frequency domain. Specifically,
evaluating a FHT for truncated messages can be made more efficient
if we neutralise all operations that apply to the constant message
tail corresponding to the truncated portion of the message.
In \cite{sayir2014b}, the expected number of operations in a FHT
on truncated messages was evaluated both exactly and using an
approximation approach that makes it easier to compute for large
alphabet sizes. The resulting comparison is promising and shows
that much can be gained in operating in the frequency domain on
truncated messages. The study however is limited to the direct
transform and stops short of treating the more difficult question
of how to efficiently evaluate the inverse transform when one 
is only interested in its $q'$ most significant output values.

\subsection{LDPC codes over rings and Analog Digital Belief Propagation (ADBP)}

Consider the problem of designing a high spectral  efficient transmission system making use of an
encoder of  rate $r_c$ and a high order $q$-PAM constellation, yielding a spectral efficiency
$\eta=r_c \log_2(q)$ [bits/dimension].

\begin{figure}
            \centering\includegraphics[angle=-90,width=\the\hsize,clip]{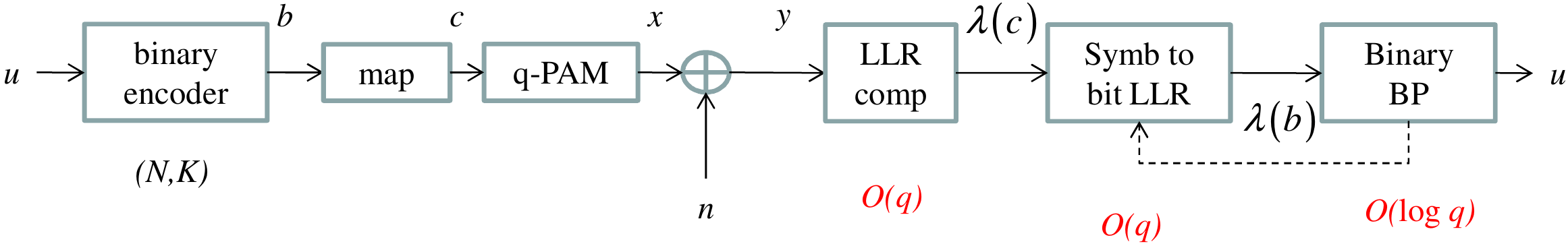}\\\vspace{.1cm}(A)
            \centering\includegraphics[angle=-90,width=\the\hsize,clip]{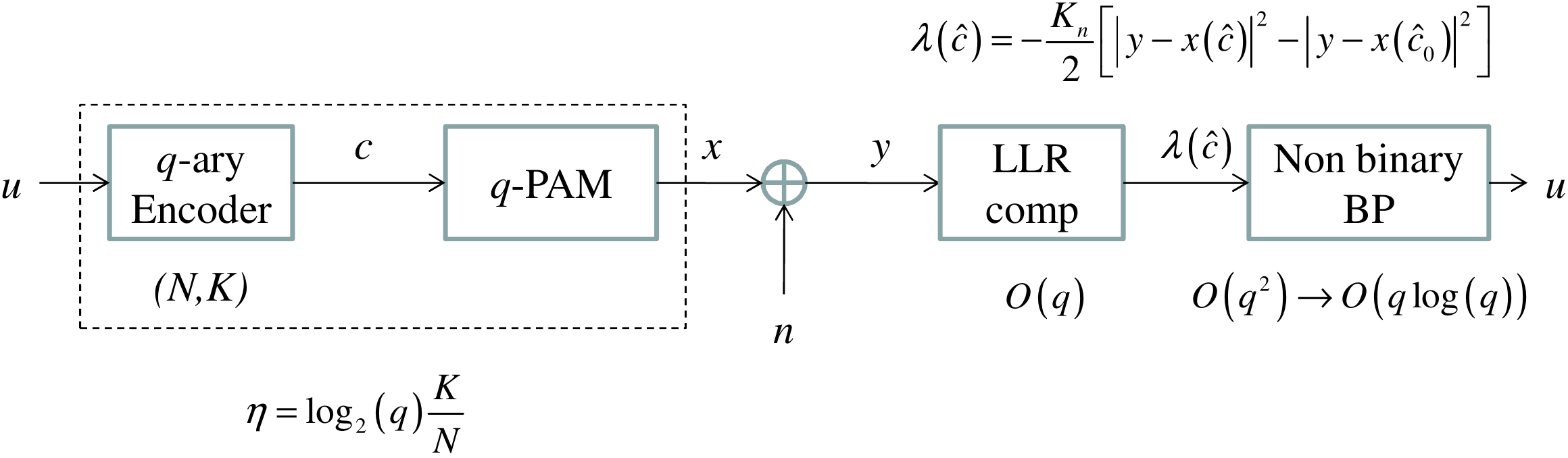}\\\vspace{.1cm}(B)
              \centering\includegraphics[angle=-90,width=\the\hsize,clip]{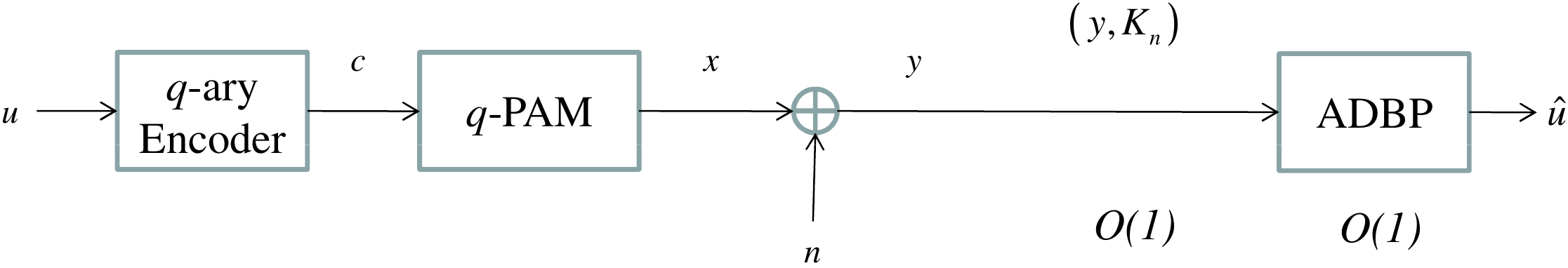}\\\vspace{.1cm} (C)
          \caption{High spectrally efficient systems using binary codes and  pragmatic receiver (A), non binary codes and non binary BP (B),  and ADBP (C).}\label{fig:BPsystem}
         \end{figure}

The current state-of the art solution, adopted in most standards, is  the \emph{pragmatic} approach
of Figure~\ref{fig:BPsystem}.(A). A \emph{binary} encoder is paired to a $q$-PAM modulation using
an interleaver and a proper mapping that produces a sequence of constellation points. At the
receiver a detector computes binary Log-Likelihood Ratios from symbol LLRs and passes them to the
binary iterative decoder through a suitably designed interleaver. The complexity of the LLR
computation is linear with $q$ and consequently exponential with the spectral efficiency $\eta$.

The feed-forward receiver scheme is associated to a ``pragmatic'' capacity that is smaller than
that of the modulation set and can be maximized using Gray mapping.
The feedback structure (dashed lines in Figure~\ref{fig:BPsystem}.(A)) can recover this capacity loss if coupled with a proper binary code
design. However,  iterating between detector and decoder  increases the receiver complexity as the conversion from bit to
symbol LLRs and viceversa is included in the loop, so that its  complexity  is multiplied by  the number of detector iterations.

A straightforward extension of an $(N,K)$ binary encoder is obtained by substituting the binary
quantities at the input of the encoder with $q$-ary symbols. Parity-check symbols are obtained by
performing  $\bmod\, q$ sums instead of $\bmod\,2$ sums in the encoding procedure. The set of
codewords is then defined as follows:
    \[
    \mathcal{C}=\{\mathbf{c}\in \mathbb{Z}_{\mathnormal q}^{\mathnormal N}: \mathbf{Hc}=\mathbf{0}\},
    \]
where the matrix elements are constrained to take only value in $\{0,\pm 1\}$. The asymptotic properties of
this class of codes were studied in \cite{bennatan2004} and \cite{erez2005ml}, where they were named ``modulo-q" or quantized coset (MQC)  codes.
Both papers showed that they achieve the random coding exponent and thus are capable of achieving capacity.

The  $q$-ary output symbols $c$ from the encoder can then be directly mapped to $q$-PAM
constellations.  At the receiver (Figure~\ref{fig:BPsystem}.(B)) the use of the regular non binary
BP iterative decoding algorithm requires to compute the Log-Likelihood ratios  of the
transmitted symbols in the form of $q-1$-ary vectors. For AWGN the LLRs take the following form
    \begin{align*}
    \lambda(\hat{c})=-\frac{K_n}{2}\left[|y-x(\hat{c})|^2-|y-x(c_0)|^2\right] \forall c\neq c_0
    \end{align*}
    where $K_n=1/\sigma_n^2$ is the \emph{concentration} of the noise.

A straightforward implementation of non binary BP results in memory
and complexity requirements of the order of $O(q)$ and $O(q^2)$
respectively. In order to reduce the complexity of non binary
decoding, several 
decoding schemes have been proposed in
recent years.
These were discussed in the previous section
and we summarize them again here.

The first straightforward simplification is obtained at check nodes by
replacing the discrete convolution of messages, having complexity
$O(q^2)$, with the product of the message Fourier transforms. The use
of FFT brings down the complexity to $O(q\log q)$. In \cite{1312606},
the authors introduce a log-domain version of this approach that has
advantages in terms of numerical stability.

Some further simplifications have been proposed in \cite{declercq2006}
with the Extended Min Sum (EMS) algorithm, where message vectors are
reduced in size by keeping only those elements in the alphabet with
higher reliability. In \cite{4392200,voicila2010} the same authors propose a
hardware implementation of the EMS decoding algorithm for non-binary
LDPC codes.

In \cite{4595129} the Min-Max algorithm is introduced with a reduced complexity architecture called
selective implementation, which can reduce by a factor 4 the operations required at the check
nodes; however, complexity is still in the order of $O(q^2)$.

Several studies on VLSI implementation of non binary decoders based on the previous algorithms have
been presented in literature
\cite{4798180,5382559,6132390,chen2012efficient,6177696,6021351,6494325}. The results of such
studies confirm that all non binary decoders require complexity growing with the size of the
alphabet.

The analog digital belief propagation (ADBP) algorithm proposed in \cite{montorsi2012analog}
represents a breakthrough in the reduction of the complexity and memory requirements with respect
to previous proposed algorithms, as for ADBP both complexity and memory requirements are
\emph{independent} of the size $q$ of the alphabet. The main simplification of ADBP is due to the
fact that messages are not stored as vector of size $q$ containing the likelihood of the discrete
variables (or equivalently their log-likelihood ratios-LLR) but rather as the two moments, or
related quantities, of some suitable predefined class of Gaussian-like distributions. ADBP can be
cast 
into the general class of expectation-propagation algorithms described by Minka
\cite{minka2001expectation}. The main contribution in \cite{montorsi2012analog} is the definition
of the suitable class of distributions for the messages relative to wrapped and discretized
variables and the derivation of the updating equations for the message parameters at the sum and
repetition operations of the Tanner graph.

A receiver system using the  Analog Digital Belief Propagation  (Figure~\ref{fig:BPsystem}.(C)),
takes then as input messages directly the pair $(K,y)$ of the concentration of the noise and the
received samples. This pair identifies  a member of the  predefined class of Gaussian-like
likelihoods and ADBP performs the BP updating by constraining the messages in the graph to stay in
the same distribution class.

The exact ADBP updating equations however are not suitable for a straightforward implementation due
to the presence of complex non linear operations. Some simplifications to the updating equations
have been presented in \cite{montorsi2012ICC}. 
In \cite{awais2014vlsi} the  practical feasibility of ADBP decoding is proved and post synthesis
results  of the hardware  implementation of required processing functions are provided.

The
ADBP decoder cannot be applied to all types of linear codes over $GF(q)$ as multiplication by field
elements different from ${\pm 1}$ is not allowed in the graph. This constraint has not been taken
into consideration previously at the code 
design stage and requires the construction of new and
efficient codes. Although  \cite{bennatan2004} and \cite{erez2005ml} show that
asymptotically this class of codes can achieve capacity, in literature there are no example of good
code constructions with finite size.

The exceptional complexity reduction achieved from using the ADBP, together with the  asymptotic
results motivates for further research effort in the design of good LDPC encoders within this
class.

\section{Polar Codes}
\label{sec:polar}

Since its inception, the major challenge in coding theory has been to
find methods that would achieve Shannon limits using low-complexity
methods for code construction, encoding, and decoding. A solution to
this problem has been proposed in \cite{ArikanIT2009} through a method
called ``channel polarization.''  Rather than attacking the coding
problem directly, the polarization approach follows a purely
information-theoretic route whereby $N$ independent identical copies
of a given binary-input channel $W$ are manipulated by certain
combining and splitting operations to ``manufacture'' a second set of
binary-input channels $\{W^{(i)}\}_{i=1}^N$ that have capacities
either near 0 or near 1, except for a fraction that vanishes as $N$
becomes large. Once such polarized channels are obtained, ``polar
coding'' consists of transmitting information at full rate over
$W^{(i)}$ that are near perfect and fixing the inputs of the remaining
channels, say, to zero.  In \cite{ArikanIT2009}, it was shown that
polar codes contructed in this manner could achieve capacity with
encoding and decoding methods of complexity $O(N\log N)$.  In
subsequent work \cite{ArikanTelatarISIT2009}, it was shown that the
probability of frame error for polar codes goes to zero roughly as
$e^{-\sqrt{N}}$ for any fixed rate below capacity; this result was later 
refined by \cite{hassani2013rate} who determined the explicit form of the dependence of
the exponent on the code rate. 

The basic binary polar code is a linear code defined for any block length $N=2^n$ in terms of a 
generator matrix 
\begin{align}\label{eq:GM}
\mathbf{G}_N & = \mathbf{F}^{\otimes n}, \quad \mathbf{F}=\begin{bmatrix} 1 & 0 \\ 1 & 1 \end{bmatrix},
\end{align}
where $\mathbf{F}^{\otimes m}$ denotes the $n$th Kronecker power of $F$.
In polar coding one encodes a data word $\mathbf{u}=(u_1,\ldots,u_N)$ into a codeword $\mathbf{x}=(x_1,\ldots,x_N)$ through the transformation 
$\mathbf{x} = \mathbf{u} \mathbf{G}_N$.
For a rate  $K/N$ polar code, one fixes $N-K$ of the coordinates of $\mathbf{u}$ to zero, effectively reducing $\mathbf{G}_N$
to a $K\times N$ matrix. For example, for a $(N,K)=(8,4)$ polar code, one may fix $u_1,u_2,u_3,u_5$ to zero and obtain from
\begin{align*}
\mathbf{G}_8 
& =
\begin{bmatrix}
1 & 0 & 0 & 0 & 0 & 0 & 0 & 0\\
1 & 1 & 0 & 0 & 0 & 0 & 0 & 0\\
1 & 0 & 1 & 0 & 0 & 0 & 0 & 0\\
1 & 1 & 1 & 1 & 0 & 0 & 0 & 0\\
1 & 0 & 0 & 0 & 1 & 0 & 0 & 0\\
1 & 1 & 0 & 0 & 1 & 1 & 0 & 0\\
1 & 0 & 1 & 0 & 1 & 0 & 1 & 0\\
1 & 1 & 1 & 1 & 1 & 1 & 1 & 1
\end{bmatrix}
\end{align*}
the $4\times 8$ generator matrix
\begin{align*}
\mathbf{G}_{4,8}
& =
\begin{bmatrix}
1 & 1 & 1 & 1 & 0 & 0 & 0 & 0\\
1 & 1 & 0 & 0 & 1 & 1 & 0 & 0\\
1 & 0 & 1 & 0 & 1 & 0 & 1 & 0\\
1 & 1 & 1 & 1 & 1 & 1 & 1 & 1
\end{bmatrix}.
\end{align*}

The 
polar code design problem consists 
in determining which
set of $(N-K)$ coordinates to freeze so as to achieve the best
possible performance under SC decoding on a given channel.  It turns
out
that 
the solution to this problem depends on the channel at hand, so in
general there is no universal set of coordinates that are guaranteed
to work well for all channels of a given capacity.  In
\cite{ArikanIT2009}, only a heuristic method was given for the polar
code design problem.  The papers \cite{MoriTanakaCL2009},
\cite{TalVardy2013}, \cite{pedarsani2011construction} provided a full
solution with complexity $O(N)$.  With this development, polar codes
became the first provably capacity-achieving class of codes with
polynomial-time algorithms for code construction, encoding, and
decoding.

Other important early theoretical contributions came in rapid succession from
\cite{hussami2009performance}, \cite{korada2010polar},
\cite{sasoglu2010polar}, \cite{korada2010empirical},
\cite{abbe2010mac}. Polar coding was extended to non-binary alphabets
in \cite{sasoglu2009polarization}, \cite{karzand2010polar},
\cite{park2013polar}, \cite{sahebi2011multilevel}. Polar code designs
by using alternative generator matrices with the goal of improving the
code performance were studied in \cite{korada2010polar2},
\cite{mori2010channel}, \cite{mori2012source},
\cite{presman2011binary}, \cite{presman2011polar}.
 
As stated above, polar coding is a channel dependent design. The performance of polar
code under ``channel mismatch'' (i.e., using a polar code optimized
for one channel on a different one) has been studied by
\cite{hassani2009compound}, who showed that there would be a rate
loss. As shown in \cite{sasoglu2011polar}, the non-universality of
polar codes is a property of the suboptimal low-complexity successive
cancellation decoding algorithm; under ML decoding, polar codes are
universal. More precisely, \cite{sasoglu2011polar} shows that a polar
code optimized for a Binary Symmetric Channel (BSC) achieves the
capacity of any other binary-input channel of the same capacity under
ML decoding. This result is very interesting theoretically since it
gives a constructive universal code for all binary-input channels;
however, it does this at the expense of giving up the $O(N\log N)$
decoding algorithm.  In more recent work \cite{sasoglu2013universal},
\cite{hassani2013universal}, universal polar coding schemes have been
described, which come at the expense of lengthening the regular polar
code construction.

It was recognized from the beginning that the finite length
performance of polar codes was not competitive with the
state-of-the-art. This was in part due to the suboptimal nature of the
standard successive cancellation (SC) decoding algorithm, and in part
due to the relatively weak minimum distance properties of these
codes. Another negative point was that the SC decoder made its
decisions sequentially, which meant that the decoder latency would
grow at least linearly with the code length, which resulted in a
throughput bottleneck. Despite these shortcomings, interest in polar
codes for potential applications continued.  The reason for this
continued interest may be attributed to several factors. First, polar
codes are firmly rooted in sound well-understood theoretical principles. 
Second, while the performance of the basic polar code is not competitive with the
state-of-the-art at short to practical block length, they are still good enough to
maintain hope that with enhancements they can become a viable alternative. This is not surprising
given that polar codes are close cousins of Reed-Muller codes, which are still an
important family of codes \cite{costello2007channel} in many respects, including performance.
Third, polar codes have the unique property that their code rate can be adjusted from 0 to 1 without
changing the encoder and decoder. Fourth, 
polar codes have a recursive structure, based on Plotkin's $|u|u+v|$
construction \cite{Plotkin}, which makes them highly suitable
for implementation in hardware. For these and
other reasons, there have been a great number of proposals in the last few years to improve
the performance of polar codes while retaining their attractive
properties. The proposed methods may be classified essentially into two categories as
encoder-side and decoder-side techniques.

Among the encoder-side techniques, one may count the non-binary polar
codes and binary polar codes starting with a larger base matrix
(kernel); however, these techniques have not yet attracted much
attention from a practical viewpoint due to their complexity. Other
encoder side techniques that have been tried include the usual
concatenation schemes with Reed-Solomon codes
\cite{bakshi2010concatenated}, and other concatenation schemes
\cite{arikan2009iscta}, \cite{mahdavifar2014performance},
\cite{trifonov2011generalized}.

Two decoder-side techniques that have been tried early on to improve
polar code performance are belief propagation (BP) decoding
\cite{ArikanCOMMLETTERS2008} and trellis-based ML decoding
\cite{ArikanICT2009}. The BP decoder did not improve the SC decoder
performance by any significant amount; however, it continues to be of
interest since the BP decoder has the potential to achieve higher
throughputs compared to SC decoding \cite{Park2014}.  

The most notable
improvement in polar coding performance came by using a
list decoder \cite{tal2011list} with CRC, which achieved near ML performance with complexity roughly $O(LN\log N)$ for a list size $L$ and code length $N$.
The CRC helps in two ways. First, it increases the code minimum distance at relatively small cost in terms of coding efficiency, thus improving code performance especially at high SNR. Second, the CRC helps select the correct codeword from the set of candidate codewords offered by the list decoder. 
It should be mentioned that the above list decoding algorithm for polar codes was an adaptation of an earlier similar algorithm given in \cite{dumer2006soft} in the context of RM codes. The vast literature on RM codes continues to be a rich source of ideas in terms of design of efficient decoding techniques for polar codes. A survey of RM codes from the perspective of decoders for polar codes has been given in \cite{Arikan2010Cairo}.
 
\begin{figure}[t]
 \centering
    \includegraphics[width=1\linewidth]{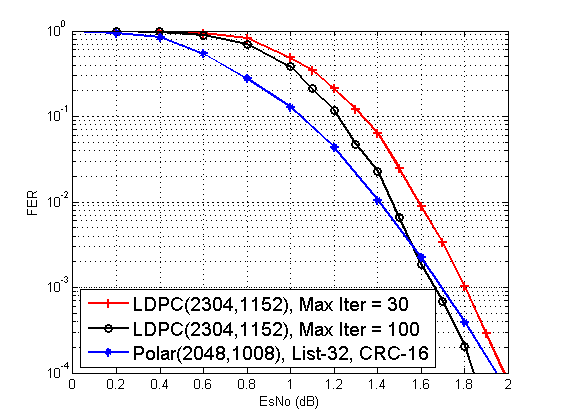}
    \caption{\label{fig:polarldpccomp} Performance comparison of polar and LDPC codes.}
\end{figure}

We end this survey by giving a performance result for polar codes. 
Figure~\ref{fig:polarldpccomp} compares the performance of a $(2048,1008)$ polar code with the WiMAX (2304,1152) LDPC code.
The polar code is obtained from a $(2048,1024)$ code by inserting a 16-bit CRC into the data and is decoded by a list-of-32 decoder.  
The LDPC code results are from the database provided by \cite{CML}; decoding is by belief propagation with maximum number of iterations limited to 30 and 100 in the results presented.
The realization that polar coding performance can rival the state-of-the-art has spurred intense research for practical implementations of these codes. 
We omit from this survey the implementation-oriented papers since that is already a very large topic by itself.
Whether polar codes will ever appear as part of the portfolio of
solutions in future systems remains uncertain. The state-of-the-art in
error correction coding is mature, with a firm footprint by turbo and
LDPC codes. Whether polar codes offer significant advantages to make
room for themselves in practical applications depends in large part on
further innovation on the subject.

\section{Conclusion}

We have presented three areas of active research in coding theory.
We introduced spatially coupled LDPC codes for which the asymptotic
performance of the iterative decoder is improved to that of the
optimal decoder. We have discussed non-binary LDPC codes and have
introduced a new decoding algorithm, analog digital belief propagation (ADBP),
whose complexity does not increase with the alphabet size. 
Finally, we have described polar coding, a novel code construction
based on a phenomenon coined channel polarization, which can be
proved theoretically to achieve channel capacity. We have stated
a number of open problems, among them:
\begin{itemize}
\item When decoding non-binary LDPC codes in the frequency domain,
  can we design a reduced complexity inverse transform if
  we are only interested in the larger components of the resulting 
  distribution-valued message?
\item How do we design LDPC codes over rings of integers to optimize
  the performance of the ADBP decoder?
\item While the potential of polar codes is established and proven,
  how can we improve the performance of its low complexity sub-optimal
  decoders at moderate codeword lengths in order for them to rival
  the performance of LDPC and turbo codes in practice? Can the performance
  of belief propagation be improved in this context, or are there 
  perhaps brand-new decoding approaches that could solve this dilemna?
\end{itemize}
We hope to have shown in this paper that coding theory is an active
area of research with many challenges remaining and a number of 
promising innovations on their way to maturing into technological
advances in the coming years.




%
\bibliography{IEEEabrv,globalrefs}

\begin{thebibliography}{10}

\bibitem{shannon1948}
C.~E. Shannon, ``A mathematical theory of communications,'' {\em Bell System
  Technical Journal}, vol.~27, pp.~379--423, July 1948.

\bibitem{golay1949}
M.~J.~E. Golay, ``Notes on digital coding,'' {\em Proceedings of the IRE},
  vol.~37, p.~657, 1949.

\bibitem{hamming1950}
R.~W. Hamming, ``Error detecing and error correcting codes,'' {\em Bell System
  Technical Journal}, vol.~29, pp.~147--160, 1950.

\bibitem{reedsolomon1960}
I.~Reed and G.~Solomon, ``Polynomial codes over certain finite fields,'' {\em
  J. Soc. Indust. Appl. Math.}, vol.~8, June 1960.

\bibitem{elias1955}
P.~Elias, ``Coding for noisy channels,'' {\em IRE Convention Record, Part 4},
  pp.~37--46, 1955.

\bibitem{viterbi1967}
A.~J. Viterbi, ``Error bounds for convolutional codes and an asymptotically
  optimum decoding algorithm,'' {\em {IEEE} Trans. Inf. Theory}, vol.~13,
  pp.~260 -- 269, Apr. 1960.

\bibitem{forney1973}
G.~Forney, ``The {V}iterbi algorithm,'' {\em Proc. {IEEE}}, vol.~61,
  pp.~258--278, Mar. 1973.

\bibitem{bcjr}
L.~Bahl, J.~Cocke, F.~Jelinek, and J.~Raviv, ``Optimal decoding of linear codes
  for minimizing symbol error rate,'' {\em {IEEE} Trans. Inf. Theory},
  pp.~284--287, Mar. 1974.

\bibitem{ungerboeck1982}
G.~Ungerboeck, ``Channel coding with multilevel/phase signals,'' {\em {IEEE}
  Trans. Inf. Theory}, pp.~55--67, Jan. 1982.

\bibitem{berrou1993}
C.~Berrou, A.~Glavieux, and P.~Thitimajshima, ``Near shannon limimt error
  correcting coding and decoding: Turbo-codes,'' in {\em Proc. {IEEE} Int.
  Conf. Commun. ({ICC})}, (Geneva, Switzerland), pp.~1064 -- 1070 vol.2, May
  1993.

\bibitem{gallager-thesis}
R.~G. Gallager, {\em Low Density Parity Check Codes}.
\newblock PhD thesis, Massachussetts Institute of Technology ({MIT}),
  Cambridge, Mass., 1963.

\bibitem{mackay1999}
D.~J.~C. MacKay, ``Good error-correcting codes based on very sparse matrices,''
  {\em {IEEE} Trans. Inf. Theory}, vol.~45, pp.~399--431, Mar. 1999.

\bibitem{massey1974}
J.~L. Massey, ``Coding and modulation in digital communications,'' in {\em
  Proc. Int. Zurich Seminar on Communications (IZS)}, (Zurich, Switzerland),
  1974.

\bibitem{ArikanIT2009}
E.~Ar{\i}kan, ``Channel polarization: A method for constructing
  capacity-achieving codes for symmetric binary-input memoryless channels,''
  {\em {IEEE} Trans. Inf. Theory}, vol.~55, pp.~3051--3073, July 2009.

\bibitem{RU01}
T.~Richardson, M.~Shokrollahi, and R.~Urbanke, ``Design of capacity-approaching
  irregular low-density parity-check codes,'' {\em {IEEE} Trans. Inf. Theory},
  vol.~47, pp.~619--637, Feb 2001.

\bibitem{JZ99}
A.~Jimenez~Felstrom and K.~Zigangirov, ``Time-varying periodic convolutional
  codes with low-density parity-check matrix,'' {\em {IEEE} Trans. Inf.
  Theory}, vol.~45, pp.~2181 --2191, Sep. 1999.

\bibitem{Tho03}
J.~Thorpe, ``Low-density parity-check {(LDPC)} codes constructed from
  protographs,'' {\em In IPN Progress Report 42-154, JPL}, Aug. 2003.

\bibitem{LFZC09}
M.~Lentmaier, G.~Fettweis, K.~Zigangirov, and D.~Costello, ``Approaching
  capacity with asymptotically regular {LDPC} codes,'' in {\em Proc. Inf.
  Theory and Applications Workshop}, (San Diego, CA, USA), pp.~173 --177, Feb.
  2009.

\bibitem{LSCZ10}
M.~Lentmaier, A.~Sridharan, D.~Costello, and K.~Zigangirov, ``Iterative
  decoding threshold analysis for {LDPC} convolutional codes,'' {\em {IEEE}
  Trans. Inf. Theory}, vol.~56, pp.~5274 --5289, Oct. 2010.

\bibitem{KRU11}
S.~Kudekar, T.~Richardson, and R.~Urbanke, ``Threshold saturation via spatial
  coupling: Why convolutional {LDPC} ensembles perform so well over the
  {BEC},'' {\em {IEEE} Trans. Inf. Theory}, vol.~57, pp.~803--834, Feb. 2011.

\bibitem{KRU12}
S.~Kudekar, T.~Richardson, and R.~Urbanke, ``Spatially coupled ensembles
  universally achieve capacity under belief propagation,'' in {\em Proc. {IEEE}
  Int. Symp. Inform. Theory ({ISIT})}, (Cambridge, MA, USA), pp.~453--457, July
  2012.

\bibitem{KYMP12}
S.~Kumar, A.~Young, N.~Maoris, and H.~Pfister, ``A proof of threshold
  saturation for spatially-coupled {LDPC} codes on {BMS} channels,'' in {\em
  Proc. Allerton Conf. on Communications, Control, and Computing},
  pp.~176--184, Oct 2012.

\bibitem{MU05}
C.~Measson and R.~Urbanke, ``Maximum a posteriori decoding and turbo codes for
  general memoryless channels,'' in {\em Proc. {IEEE} Int. Symp. Inform. Theory
  ({ISIT})}, (Adelaide, Australia), pp.~1241--1245, Sept 2005.

\bibitem{MLC10}
D.~Mitchell, M.~Lentmaier, and D.~Costello, ``New families of {LDPC} block
  codes formed by terminating irregular protograph-based {LDPC} convolutional
  codes,'' in {\em Proc. {IEEE} Int. Symp. Inform. Theory ({ISIT})}, (Austin,
  Texas, USA), pp.~824--828, June 2010.

\bibitem{NLF14}
W.~Nitzold, G.~Fettweis, and M.~Lentmaier, ``Spatially-coupled nearly-regular
  {LDPC} code ensembles for rate-flexible code design,'' in {\em Proc. {IEEE}
  Int. Conf. Commun. ({ICC})}, (Sydney, Australia), pp.~2027--2032, June 2014.

\bibitem{Iyengar2012}
A.~R. Iyengar, M.~Papaleo, P.~H. Siegel, J.~K. Wolf, A.~Vanelli-Coralli, and
  G.~E. Corazza, ``Windowed decoding of protograph-based {LDPC} convolutional
  codes over erasure channels,'' {\em {IEEE} Trans. Inf. Theory}, vol.~58,
  pp.~2303--2320, Apr. 2012.

\bibitem{HLF+12}
N.~Ul~Hassan, M.~Lentmaier, and G.~Fettweis, ``Comparison of {LDPC} block and
  {LDPC} convolutional codes based on their decoding latency,'' in {\em Proc.
  Int. Symp. on Turbo Codes \& Iterative Inf. Proc.}, (Gothenburg, Sweden),
  Aug. 2012.

\bibitem{HH09}
T.~Hehn and J.~Huber, ``{LDPC} codes and convolutional codes with equal
  structural delay: a comparison,'' {\em {IEEE} Trans. Commun.}, vol.~57,
  pp.~1683--1692, June 2009.

\bibitem{MCFF10}
S.~Maiya, D.~Costello, T.~Fuja, and W.~Fong, ``Coding with a latency
  constraint: The benefits of sequential decoding,'' in {\em Proc. Allerton
  Conf. on Communications, Control, and Computing}, pp.~201 --207, Oct. 2010.

\bibitem{SPL09}
E.~Sharon, N.~Presman, and S.~Litsyn, ``Convergence analysis of generalized
  serial message-passing schedules,'' {\em {IEEE} J. Sel. Areas Commun.},
  vol.~27, pp.~1013 --1024, Aug. 2009.

\bibitem{HPLFC12}
N.~Ul~Hassan, A.~Pusane, M.~Lentmaier, G.~Fettweis, and J.~Costello, D.J.,
  ``Reduced complexity window decoding schedules for coupled {LDPC} codes,'' in
  {\em Proc. {IEEE} Inform. Theory Workshop ({ITW})}, (Lausanne, Switzerland),
  pp.~20 -- 24, Sep. 2012.

\bibitem{HPLFC13}
N.~Ul~Hassan, A.~Pusane, M.~Lentmaier, G.~Fettweis, and J.~Costello, D.J.,
  ``Non-uniform windowed decoding schedules for spatially coupled codes,'' in
  {\em Proc. {IEEE} Glob. Comm. Conf. ({GLOBECOM})}, (Atlanta, GA, USA),
  pp.~1862--1867, Dec 2013.

\bibitem{OSW94}
L.~Ozarow, S.~Shamai~(Shitz), and A.~Wyner, ``Information theoretic
  considerations for cellular mobile radio,'' {\em {IEEE} Trans. Veh.
  Technol.}, vol.~43, no.~2, pp.~359--378, 1994.

\bibitem{BCT00}
E.~Biglieri, G.~Caire, and G.~Taricco, ``Coding for the fading channel: a
  survey,'' {\em Signal Processing (EURASIP)}, vol.~80, no.~7, pp.~1135--1148,
  2000.

\bibitem{BGBZ10}
J.~Boutros, A.~Guill{\'e}n~i F{\`a}bregas, E.~Biglieri, and G.~Zemor,
  ``Low-density parity-check codes for nonergodic block-fading channels,'' {\em
  {IEEE} Trans. Inf. Theory}, vol.~56, no.~9, pp.~4286--4300, 2010.

\bibitem{HLAF14}
N.~Ul~Hassan, M.~Lentmaier, I.~Andriyanova, and G.~Fettweis, ``Improving code
  diversity on block-fading channels by spatial coupling,'' in {\em Proc.
  {IEEE} Int. Symp. Inform. Theory ({ISIT})}, (Honolulu, Hawaii, USA),
  pp.~2311--2315, June 2014.

\bibitem{davey1998}
M.~C. Davey and D.~J.~C. MacKay, ``Low density parity check codes over
  {GF(q)},'' in {\em Proc. {IEEE} Inform. Theory Workshop ({ITW})}, (Killarney,
  Ireland), pp.~70--71, June 1998.

\bibitem{declercq2006}
D.~Declercq and M.~P. Fossorier, ``Decoding algorithms for nonbinary {LDPC}
  codes over {GF(q)},'' {\em {IEEE} Trans. Commun.}, vol.~55, pp.~633--643,
  Apr. 2007.

\bibitem{voicila2010}
A.~Voicila, D.~Declercq, F.~Verdier, M.~Fossorier, and P.~Urard,
  ``Low-complexity decoding for non-binary {LDPC} codes in high order fields,''
  {\em {IEEE} Trans. Commun.}, pp.~1365--1375, May 2010.

\bibitem{boutillon2010}
E.~Boutillon and L.~Conde-Canencia, ``Bubble check: a simplified algorithm for
  elementary check node processing in extended min-sum non-binary {LDPC}
  decoders,'' {\em {IET} Electronic Letters}, vol.~46, pp.~633--634, Apr. 2010.

\bibitem{montorsi2012analog}
G.~Montorsi, ``Analog digital belief propagation,'' {\em {IEEE} Commun. Lett.},
  vol.~16, no.~7, pp.~1106--1109, 2012.

\bibitem{sayir2014b}
J.~Sayir, ``Non-binary {LDPC} decoding using truncated messages in the
  walsh-hadamard domain,'' in {\em Proc. Int. Symp. Inf. Theory and Its App.
  ({ISITA})}, (Melbourne, Australia), Oct. 2014.

\bibitem{bennatan2004}
A.~Bennatan and D.~Burshtein, ``On the application of {LDPC} codes to arbitrary
  discrete memoryless channels,'' {\em {IEEE} Trans. Inf. Theory}, vol.~50,
  pp.~417--438, Mar. 2004.

\bibitem{erez2005ml}
U.~Erez and G.~Miller, ``{The ML decoding performance of LDPC ensembles over
  $Z_q$},'' {\em {IEEE} Trans. Inf. Theory}, vol.~51, no.~5, pp.~1871--1879,
  2005.

\bibitem{1312606}
H.~Wymeersch, H.~Steendam, and M.~Moeneclaey, ``{Log-domain decoding of LDPC
  codes over GF($q$)},'' in {\em Proc. {IEEE} Int. Conf. Commun. ({ICC})},
  vol.~2, (Paris, France), pp.~772--776 Vol.2, 2004.

\bibitem{4392200}
A.~Voicila, F.~Verdier, D.~Declercq, M.~Fossorier, and P.~Urard,
  ``{Architecture of a low-complexity non-binary LDPC decoder for high order
  fields},'' in {\em Communications and Information Technologies, 2007. ISCIT
  '07. International Symposium on}, pp.~1201--1206, 2007.

\bibitem{4595129}
V.~Savin, ``{Min-Max decoding for non binary LDPC codes},'' in {\em Proc.
  {IEEE} Int. Symp. Inform. Theory ({ISIT})}, (Toronto, Canada), pp.~960--964,
  2008.

\bibitem{4798180}
C.~Spagnol, E.~Popovici, and W.~Marnane, ``{Hardware Implementation of GF($q$)
  {LDPC} Decoders},'' {\em {IEEE} Trans. Circuits Syst. {I}}, vol.~56, no.~12,
  pp.~2609--2620, 2009.

\bibitem{5382559}
J.~Lin, J.~Sha, Z.~Wang, and L.~Li, ``{An Efficient VLSI Architecture for
  Nonbinary LDPC Decoders},'' {\em {IEEE} Trans. Circuits Syst. {II}}, vol.~57,
  no.~1, pp.~51--55, 2010.

\bibitem{6132390}
C.~Zhang and K.~Parhi, ``{A Network-Efficient Nonbinary QC-LDPC Decoder
  Architecture},'' {\em {IEEE} Trans. Circuits Syst. {I}}, vol.~59, no.~6,
  pp.~1359--1371, 2012.

\bibitem{chen2012efficient}
X.~Chen, S.~Lin, and V.~Akella, ``{Efficient configurable decoder architecture
  for nonbinary Quasi-Cyclic LDPC codes},'' {\em {IEEE} Trans. Circuits Syst.
  {I}}, vol.~59, no.~1, pp.~188--197, 2012.

\bibitem{6177696}
X.~Chen and C.-L. Wang, ``{High-Throughput Efficient Non-Binary LDPC Decoder
  Based on the Simplified Min-Sum Algorithm},'' {\em {IEEE} Trans. Circuits
  Syst. {I}}, vol.~59, no.~11, pp.~2784--2794, 2012.

\bibitem{6021351}
Y.-L. Ueng, C.-Y. Leong, C.-J. Yang, C.-C. Cheng, K.-H. Liao, and S.-W. Chen,
  ``An efficient layered decoding architecture for nonbinary {QC-LDPC} codes,''
  {\em {IEEE} Trans. Circuits Syst. {I}}, vol.~59, no.~2, pp.~385--398, 2012.

\bibitem{6494325}
Y.-L. Ueng, K.-H. Liao, H.-C. Chou, and C.-J. Yang, ``A high-throughput
  trellis-based layered decoding architecture for non-binary {LDPC} codes using
  {Max-Log-{QSPA}},'' {\em {IEEE} Trans. Signal Process.}, vol.~61, no.~11,
  pp.~2940--2951, 2013.

\bibitem{minka2001expectation}
T.~P. Minka, ``Expectation propagation for approximate bayesian inference,'' in
  {\em Proceedings of the Seventeenth conference on Uncertainty in artificial
  intelligence}, pp.~362--369, Morgan Kaufmann Publishers Inc., 2001.

\bibitem{montorsi2012ICC}
G.~Montorsi, ``Analog digital belief propagation: from theory to practice,'' in
  {\em Proc. {IEEE} Int. Conf. Commun. ({ICC})}, pp.~2591--2595, IEEE, 2012.

\bibitem{awais2014vlsi}
M.~Awais~Aslam, G.~Masera, M.~Martina, and G.~Montorsi, ``{VLSI} implementation
  of a non-binary decoder based on the analog digital belief propagation,''
  {\em {IEEE} Trans. Signal Process.}, pp.~3965--3975, July 2014.

\bibitem{ArikanTelatarISIT2009}
E.~Ar{\i}kan and E.~Telatar, ``On the rate of channel polarization,'' in {\em
  Proc. {IEEE} Int. Symp. Inform. Theory ({ISIT})}, (Seoul, South Korea),
  pp.~1493--1495, 28 June - July 3 2009 2009.

\bibitem{hassani2013rate}
S.~H. Hassani, R.~Mori, T.~Tanaka, and R.~L. Urbanke, ``Rate-dependent analysis
  of the asymptotic behavior of channel polarization,'' {\em {IEEE} Trans. Inf.
  Theory}, vol.~59, no.~4, pp.~2267--2276, 2013.

\bibitem{MoriTanakaCL2009}
R.~Mori and T.~Tanaka, ``Performance of polar codes with the construction using
  density evolution,'' {\em {IEEE} Commun. Lett.}, vol.~13, pp.~519--521, July
  2009.

\bibitem{TalVardy2013}
I.~Tal and A.~Vardy, ``How to construct polar codes,'' {\em {IEEE} Trans. Inf.
  Theory}, vol.~59, pp.~6562--6582, Oct 2013.

\bibitem{pedarsani2011construction}
R.~Pedarsani, S.~H. Hassani, I.~Tal, and I.~Telatar, ``On the construction of
  polar codes,'' in {\em Proc. {IEEE} Int. Symp. Inform. Theory ({ISIT})},
  (St.~Petersburg, Russia), pp.~11--15, IEEE, 2011.

\bibitem{hussami2009performance}
N.~Hussami, S.~B. Korada, and R.~Urbanke, ``Performance of polar codes for
  channel and source coding,'' in {\em Proc. {IEEE} Int. Symp. Inform. Theory
  ({ISIT})}, (Seoul, South Korea), pp.~1488--1492, IEEE, 2009.

\bibitem{korada2010polar}
S.~B. Korada and R.~Urbanke, ``Polar codes for slepian-wolf, wyner-ziv, and
  gelfand-pinsker,'' in {\em Proc. {IEEE} Inform. Theory Workshop ({ITW})},
  (Cairo, Egypt), 2010.

\bibitem{sasoglu2010polar}
E.~Sasoglu, E.~Telatar, and E.~Yeh, ``Polar codes for the two-user binary-input
  multiple-access channel,'' in {\em Proc. {IEEE} Inform. Theory Workshop
  ({ITW})}, (Cairo, Egypt), pp.~1--5, IEEE, 2010.

\bibitem{korada2010empirical}
S.~B. Korada, A.~Montanari, I.~Telatar, and R.~Urbanke, ``An empirical scaling
  law for polar codes,'' in {\em Proc. {IEEE} Int. Symp. Inform. Theory
  ({ISIT})}, (Austin, Texas, USA), pp.~884--888, IEEE, 2010.

\bibitem{abbe2010mac}
E.~Abbe and I.~Telatar, ``Mac polar codes and matroids,'' in {\em Proc. Inf.
  Theory and Applications Workshop}, (San Diego, CA, USA), pp.~1--8, IEEE,
  2010.

\bibitem{sasoglu2009polarization}
E.~Sasoglu, I.~Telatar, and E.~Ar{\i}kan, ``Polarization for arbitrary discrete
  memoryless channels,'' in {\em Proc. {IEEE} Inform. Theory Workshop ({ITW})},
  (Taormina, Italy), pp.~144--148, IEEE, 2009.

\bibitem{karzand2010polar}
M.~Karzand and I.~Telatar, ``Polar codes for $q$-ary source coding,'' in {\em
  Proc. {IEEE} Int. Symp. Inform. Theory ({ISIT})}, (Austin, TX, USA),
  pp.~909--912, IEEE, 2010.

\bibitem{park2013polar}
W.~Park and A.~Barg, ``Polar codes for $q$-ary channels,'' {\em {IEEE} Trans.
  Inf. Theory}, vol.~59, no.~2, pp.~955--969, 2013.

\bibitem{sahebi2011multilevel}
A.~G. Sahebi and S.~S. Pradhan, ``Multilevel polarization of polar codes over
  arbitrary discrete memoryless channels,'' in {\em Proc. Allerton Conf. on
  Communications, Control, and Computing}, pp.~1718--1725, IEEE, 2011.

\bibitem{korada2010polar2}
S.~B. Korada, E.~Sasoglu, and R.~Urbanke, ``Polar codes: Characterization of
  exponent, bounds, and constructions,'' {\em {IEEE} Trans. Inf. Theory},
  vol.~56, no.~12, pp.~6253--6264, 2010.

\bibitem{mori2010channel}
R.~Mori and T.~Tanaka, ``Channel polarization on q-ary discrete memoryless
  channels by arbitrary kernels,'' in {\em Proc. {IEEE} Int. Symp. Inform.
  Theory ({ISIT})}, (Austin, TX, USA), pp.~894--898, IEEE, 2010.

\bibitem{mori2012source}
R.~Mori and T.~Tanaka, ``Source and channel polarization over finite fields and
  reed-solomon matrix,'' {\em arXiv preprint arXiv:1211.5264}, 2012.

\bibitem{presman2011binary}
N.~Presman, O.~Shapira, and S.~Litsyn, ``Binary polar code kernels from code
  decompositions,'' in {\em Proc. {IEEE} Int. Symp. Inform. Theory ({ISIT})},
  (St.~Petersburg, Russia), pp.~179--183, IEEE, 2011.

\bibitem{presman2011polar}
N.~Presman, O.~Shapira, and S.~Litsyn, ``Polar codes with mixed kernels,'' in
  {\em Proc. {IEEE} Int. Symp. Inform. Theory ({ISIT})}, (St.~Petersburg,
  Russia), pp.~6--10, IEEE, 2011.

\bibitem{hassani2009compound}
S.~H. Hassani, S.~B. Korada, and R.~Urbanke, ``The compound capacity of polar
  codes,'' {\em arXiv preprint arXiv:0907.3291}, 2009.

\bibitem{sasoglu2011polar}
E.~Sasoglu, {\em Polar coding theorems for discrete systems}.
\newblock PhD thesis, EPFL, 2011.

\bibitem{sasoglu2013universal}
E.~Sasoglu and L.~Wang, ``Universal polarization,'' {\em arXiv preprint
  arXiv:1307.7495}, 2013.

\bibitem{hassani2013universal}
S.~H. Hassani and R.~Urbanke, ``Universal polar codes,'' {\em arXiv preprint
  arXiv:1307.7223}, 2013.

\bibitem{costello2007channel}
D.~J. Costello and G.~D. Forney~Jr, ``Channel coding: The road to channel
  capacity,'' {\em Proc. {IEEE}}, vol.~95, no.~6, pp.~1150--1177, 2007.

\bibitem{Plotkin}
M.~Plotkin, ``Binary codes with specified minimum distance,'' {\em {IRE Trans.
  Inform. Theory}}, vol.~6, pp.~445--450, Sept. 1960.

\bibitem{bakshi2010concatenated}
M.~Bakshi, S.~Jaggi, and M.~Effros, ``Concatenated polar codes,'' in {\em Proc.
  {IEEE} Int. Symp. Inform. Theory ({ISIT})}, (Austin, TX, USA), pp.~918--922,
  IEEE, 2010.

\bibitem{arikan2009iscta}
E.~Ar{\i}kan and G.~Markarian, ``{Two-Dimensional Polar Coding},'' in {\em
  Proc. Int. Symp. on Commun. Theory and App. (ISCTA)}, (Ambleside, U.K.), July
  2009.

\bibitem{mahdavifar2014performance}
H.~Mahdavifar, M.~El-Khamy, J.~Lee, and I.~Kang, ``Performance limits and
  practical decoding of interleaved reed-solomon polar concatenated codes,''
  {\em {IEEE} Trans. Commun.}, vol.~62, no.~5, pp.~1406--1417, 2014.

\bibitem{trifonov2011generalized}
P.~Trifonov and P.~Semenov, ``Generalized concatenated codes based on polar
  codes,'' in {\em Proc. Int. Symp. on Wireless Comm. Systems (ISWCS)},
  (Aachen, Germany), pp.~442--446, 2011.

\bibitem{ArikanCOMMLETTERS2008}
E.~Ar{\i}kan, ``A performance comparison of polar codes and {R}eed-{M}uller
  codes,'' {\em {IEEE} Commun. Lett.}, vol.~12, pp.~447--449, June 2008.

\bibitem{ArikanICT2009}
E.~Ar{\i}kan, H.~Kim, G.~Markarian, {\" U}.~{\" O}zg{\" u}r, and E.~Poyraz,
  ``Performance of short polar codes under {ML} decoding,'' in {\em Proc. ICT
  MobileSummit}, (Santander, Spain), 10-12 June 2009.

\bibitem{Park2014}
Y.~S. Park, Y.~Tao, S.~Sun, and Z.~Zhang, ``A 4.68gb/s belief propagation polar
  decoder with bit-splitting register file,'' in {\em VLSI Circuits Digest of
  Technical Papers, 2014 Symposium on}, pp.~1--2, June 2014.

\bibitem{tal2011list}
I.~Tal and A.~Vardy, ``List decoding of polar codes,'' in {\em Proc. {IEEE}
  Int. Symp. Inform. Theory ({ISIT})}, (St.~Petersburg, Russia), pp.~1--5,
  IEEE, 2011.

\bibitem{dumer2006soft}
I.~Dumer and K.~Shabunov, ``Soft-decision decoding of reed-muller codes:
  recursive lists,'' {\em {IEEE} Trans. Inf. Theory}, vol.~52, no.~3,
  pp.~1260--1266, 2006.

\bibitem{Arikan2010Cairo}
E.~Ar{\i}kan, ``A survey of reed-muller codes from polar coding perspective,''
  in {\em Proc. {IEEE} Inform. Theory Workshop ({ITW})}, (Cairo, Egypt),
  pp.~1--5, Jan 2010.

\bibitem{CML}
``{The Coded Modulation Library}.''
  http://www.iterativesolutions.com/\-Matlab.htm.

\end{thebibliography}


\end{document}